# Possibilities and challenges of STEAM pedagogies

Iván Sánchez (ivan.sanchez@oulu.fi)
Marta Cortés (marta.cortes@oulu.fi)

## STEAM concept and background

The term STEM refers to teaching and learning in the fields of science, technology, engineering, and mathematics. It is a very broad concept that includes education across all grade levels: starting from kindergarten and finishing in post-doctorate studies. STEM education is also transversal: is applied in formal (e.g. classrooms), non-formal (e.g. in courses and workshops taught at afterschool programs) and informal (e.g. courses taught in makerspaces) education environments (Gonzalez and Kuenzi, 2012). STEM education aims to help the next generation of students to solve real-world problems by utilizing knowledge of multiple disciplines and horizontal competences such as critical thinking, collaboration, and creativity (Burrows and Slater 2015).

A report issued by the National Commission on Excellence in Education (1983) aimed to describe which was the status in education in the US. In conclusion, the report included some recommendations to educate future citizens that could cope with the requirements of future jobs, many of which were not yet known. This set of recommendations has evolved in the last decades, and have been later identified globally as 21st century skills. They partly switched the education paradigm from content only to skills (or competences) and content. Such skills (Chu 2017), include examination, inquiry, critical thinking and reasoning, team work, motivation…

The term STEM was coined by the NSF (National Science Foundation) in the USA in the beginning of 2000's with the goal of attracting and training students in the fields of science and technology, in which it was expected a shortage of workforce. From that moment on, several programs were started in the USA, with the goal of introducing in-service and pre-service teachers the foundations, pedagogies, curriculum, research, and contemporary issues of the STEM education disciplines. The NSF remarked an integrative view of the disciplines. They should not be taught separately but as a whole, and thus bringing together the technological design methodologies used traditionally in technology and engineering with the inquiry learning approaches used for ages in science and math. (Sanders, 2009). The NSF expected engage students or teams of students in scientific inquiry situated in the context of technological problem-solving. For instance, instead of teaching concepts about energy, sustainability and technology in separate niches, students might build their own wind turbine, in order to acquire similar knowledge.

Land (2013) defends that teaching technology and science just prepare students to execute tasks fluidly. However, other important competences that powers innovation and creativity are left out. Teaching science and technology should come hand in hand with the development of creative thinking skills through art and design(Madden 2013). In addition, the incorporation of arts may develop further majority of STEM skills: arts utilize a divergent approach while science, technology and engineering are more convergent. Convergent thinking consists of following a process to reach one single solution to a problem, whereas divergent thinking involves exploring many possible solutions to the same problem. In that way, students will be able to simultaneous decompose a complex problem using convergent thinking and then apply the corresponding solution to the real world uses divergent thinking.The addition of the artistic skills to the science and technology education gave birth to a new acronym: STEAM (notice the addition of A for arts). The literature presents John Maeda, President of the Rhode Island School of Design, as the father of the STEAM movement (Maeda, 2012).

## STEAM pedagogies

Moore et al. (2004), after an extensive review of literature, tried to characterized STEM education. They included six main principles for quality STEM education: (1) multidisciplinar content including math, science and technology; (2) student-centered pedagogy; (3) lessons are situated in a inviting and engaging context; (4) inclusion of engineering design or redesign challenge; (5) learn by making errors; (6) teamwork and collaboration. In this section we would like to discuss a bit about the pedagogical methodologies that facilitate most of the tenets defined by Moore and colleagues.

As stated in Chu (2017) "One limitation of the twenty-first century skills models is that while they specify prioritized learning objectives, they do not offer educators the "means" by which to achieve those articulated "ends."" And this is where STEAM education quicks in. STEAM practices support switching from traditional lecture-based teaching to inquiry and project-based strategies including collaborative learning. The aim of this is to educate citizens that (Quigley and Herro 2016) approach problem solving through innovation, creativity, critical thinking, effective communication, collaboration and, of course,  new knowledge. Some  (Yakman 2008, Watson and Watson 2013)  have pointed out that these concepts were already acknowledged more than a century ago by educational philosophers such as Friedrich Fröbel or John Dewey. Fröbel promoted real world experiences and contextual learning as the only way to promote real understanding. Dewey proposed the integration of education across subjects, and to expose students to real-world applications to increase engagement, promoting collaborative building of knowledge. Both of them predated somehow constructivism. Both of them promoted student-centered learning.

When reviewing STEAM education literature, main themes in relation with pedagogy (Yakman 2008, Connor et al. 2015, Quigley and Herro. 2016) are discipline integration and several instructional approaches based on constructivist theories like problem-based learning, project-based learning and inquiry based learning.

Discipline integration (or Integrative education) implies that the solution of complex problems involve the knowledge, processes and skills from different disciplines. Benefits of discipline integration include (Stohlmann et al. 2012, Shernoff et al. 2017): more relevant, less fragmented, and more stimulating experiences for learners; its student centered approach; improvement in higher level thinking skills, problem solving, and retention; improvement on student attitudes and interest in school, their motivation to learn and achievement. Connor et al (2015) find Integrative STEM education has a strong basis in constructionism and in cognitive science. In fact, they point that the set of following cognitive themes identified by Bruning (2004) resonate with integrative STEM education:

- Learning is a constructive, not a receptive, process.
- Motivation and beliefs are integral to cognition.
- Social interaction is fundamental to cognitive development.
- Knowledge, strategies, and expertise are contextual

Connor et al. (2015) also point that there is growing evidence that integrative instruction enhances learning: Hartzler (2000) conducted a meta-analysis across 30 individual studies of the effects of integrated instruction on student achievement.

The instructional approaches involved in STEAM pedagogies that we have mainly identified are project based learning (Land 2013, Madden 2013, Watson and Watson 2013), problem based learning (Quigley and Herro 2016) and inquiry based learning ( Madden 2013). The three of them are mainly learner centered and based on the same principle: student needs to solve a problem (or answer a question). Though the boundaries among those methods are quite blurred, there are some differences, mainly on the framing of the problem and how much information on how to get to the solution is given.

Project based learning (Knoll 1997, Becket 2002) involves students creating knowledge in order to solve problems that arise while they are engaged in purposeful real-world activities. It was first presented as The Project Method by Dewey's student, Kilpatrick. Main characteristics for Project based learning are: real-world activities frame problems that can be solved along the project with no predefined steps, often with an interdisciplinary approach. However, Problem based learning (Savery and Duffy 1995) focus is in solving a problem framed in a fictional scenario that does not necessarily imply an interdisciplinary approach. The students generate the learning issues (objectives) based on their analysis of the problem. While these two imply exploring to obtain an answer, Inquiry based learning (Colburn 2000) aims to discover an answer from an initial question by formulating new questions. With inquiry students in which they develop an understanding of how scientists study the natural world. All of these instructional approaches favor collaborative learning and team work.

# Perceptions and challenges associated to STEAM education for in service teachers

Sanders (2009) remarks the important role of primary education teachers in STEAM education: students start loosing interest in science, maths and technology at an early age. When interest for science and technology decays, it is difficult to get it back.

The literature presents two main challenges teachers have to face to implement STEAM education in their classes: Integrative approach (different subjects content must be applied simultaneously) and multidisciplinary education (teachers have to teach content outside their comfort area or speciality).

In-service teachers that are currently applying STEAM education in their classes perceived the following challenges when applying integrated STEAM approaches (Shernoff et al. 2017): lack of time for collaborative planning, lack of time for instruction, an inadequate school structure and organization (e.g. scheduling), difficulty to assess STEAM achievements, lack of resources, and inadequate teacher education (pre-service education should put more focus in interdisciplinarity while in-service education should include focus on content areas outside teacher's specialization). An extensive literature review on teachers perception of STEM recently published by Margot and Kettle (2019) confirms the results reported by Shernoff and colleagues (2017). In addition, Margot and Kettle (2019) remarks how they would like to receive support: collaboration among colleagues, the availability of quality curriculum, support from learning officials and local administration, and more important, a well-organized and frequently available professional learning opportunities Other research work reporting perception of STEAM education by teachers can be found from Park et al. ( 2016) and El-Deghaidy and Mansour (2015). In both cases, conclusions align with those exposed previously.

It is clear that school teachers need proper training to integrate STEAM education in their classes. In order to be efficient dealing with students misconceptions in STEAM subjects, they must hold an thorough knowledge of STEAM concepts (Nadelson et al., 2013). This research claims that to overcome the limitations associated with the minimal preparation of in service teachers in STEAM education, it is necessary that they participate in continuing education. On the other hand, Sanders (2009) remarks that due to the integrative character of STEAM, education teachers cannot be prepared anymore in conventional ways to teach in the conventional settings. STEAM education promotes pupils engagement in inquiry, authentic applications, and active learning environments which require teachers to seek a proper pedagogic training.

In order to cope with the challenge of supporting teachers when building multidisciplinary and integrative education learning activities, literature proposes different frameworks for teacher training. Burrows and Slater (2015) defend that currently there is a disconnect between the traditional preparation (where teachers focus in one or two disciplines) and what is needed to apply STEAM education. They propose a progressive education in which teachers move from a Level Zero (single discipline) to a Level Four (Constant STEAM education). Moving between these two end points exists a testable trajectory. First step would be adding quantitative reasoning or mathematics to concepts being taught in the target discipline. After that, teachers should integrate two or more distinct science disciplines. Following step would be to add Engineering projects: materialization of concepts and ideas that came up from scientific disciplines (but emphasizing principles of engineering instead of scientific methods). When mastered these 3 levels it would be straightforward to move to the pure STEAM education.

## STEAM in undergraduate education

Despite STEAM education being conceived to cover all education spectrum: from primary to undergraduate education, main research stream has focus on school education. We could not find that much research on how STEAM education has been used in undergraduate studies. In this essay, we have omitted all papers that describe small interventions in particular courses or who are very topic specific (e.g teaching programming or robotics). We tried to choose only papers that present long-term interventions (e.g. modifying a whole course structure).

Some studies like the one presented by Brenier et al. (2012) presents the lack of awareness of the STEM or STEAM concept among the university community. In addition, an extensive literature review carried out by Henderson and colleagues (2011), analyzing 190 papers published from 1995 to 2008 shows that the researchers do not provide strong evidence of success of the tested strategies and that the different research communities are isolated.

In spite of the fact that student centric practices developed further problem solving skills, Connor et al., (2015) defend that there has been little adoption of such practices (e.g. project based learning) in undergraduate engineering education. The "chalk-and-talk" is by far the most utilized methodology in university education. In addition, they claim that there is a misconception between Problem Based Learning and Project Based Learning. In the last one, students should set up goals and outcomes, producing more creative solutions. Authors also remark one of the big problems when trying to apply student centric approach in undergraduate studies is the so called "*disciplinary egocentrism*", described previously by Ritcher and Paretti (2009). Disciplinary egocentrism is a failure to see connections between a given discipline and an interdisciplinary subject or problem and also a failure to recognize differences in perspectives and contributions. Connor and colleagues present several experiences where they try to integrate arts based pedagogics, which emphasize inductive vs. deductive teaching methods, to promote inquiry guided learning in engineering undergraduate courses. They argue that different pedagogical methods such as problem based, project based, inquiry based and discovery learning can coexist in the same instructional model. Finally, based on their experience they propose, the following set of guidelines: embrace different disciplines, design flexible projects for students, tease out creativity, allow and encourage failure, realize students are different, consider vertical orientation (mixing students from different ages groups) and explore horizontal blurring (projects and assessments to stretch over multiple courses).

Majority of efforts made to integrate STEAM education in university studies consist of changing teaching methodologies in existing undergraduate courses. For instance, Ifenthaler et al. (2015) presents how they built the undergraduate course "Designing for Open Innovation" using STEAM methodologies. It is a multidisciplinary course that includes topics from Economy, Social Sciences and Environment. The goal of the course is to answer a "question of the semester". Students divided in teams complete different assignments that 1) Allow them to identify the competences they would like to develop further 2) help them in acquiring collaboration skills and 3) facilitate the acquisition of the competences defined in the first bullet. Finally the course ended with a reflective essay. One of the main differences between this course and the old version of the course is how the assignments were used to scaffold student learning and team formation. In the presented course implementation, learning was achieved at three levels: individual learning, team learning, and learning from each other. The first group tasks were aimed to identify their own mental models. As the course advances the group assignment is aimed more to acquire the desired competences and answer the question of the semester.

All around the world there has also been some efforts to add STEAM education in undergraduate programs. For instance, in their article Madden et al. (2013) describe how they build from scratch a multidisciplinar STEAM undergraduate curriculum including biology, computer science, mathematics, music education, psychology, theatre and visual arts. While the STEM disciplines of the curriculum aims to develop more analytical skills, the artistic disciplines aimed to develop creativity and divergent thinking in students. The program goal is to develop creative leaders with the following six characteristics: good communicators, good organizers, managers with motivational skills, creative and innovative and cross-disciplinary knowledge in at least two fields. With the help of advisors students personalize their major (Student-Initiated Integrative Major) which must include courses from at least two disciplines. The students and faculty work together after the first semester in building the Integrated Learning Module. This module involves multidisciplinary problem solving teams working on a particular thematic problem. Students propose the themes and then select related problems to study, so they take ownership over their own projects. Faculty will model problem solving and encourage reflection, communication skills, and self-monitoring. Finally, students need to do internships in local companies. According to them, this is the first STEAM program taught for undergraduate education in the world.

## Discussion

It is clear that, for future professionals, it is not only important to acquire knowledge and expertise in particular fields but also the procurement of transversal competences such as creativity, critical thinking, communication, collaboration or problem solving. Many companies are looking for these skills among their job applicants. In addition, multidisciplinary / interdisciplinary is a reality that many of us have to face in our jobs. It is rather common that companies look for versatile workers which are able to work in different fields of expertise. STEM or STEAM education tries to prepare students for this new reality. It is clear that STEM education can help the next generation of students to solve real-world problems by applying concepts that cut across disciplines as well as capacities of critical thinking, collaboration, and creativity (Burrows and Slater 2015)

Although some researchers and practitioners consider STEM or STEAM just as a buzzword that is utilized with marketing purposes and to attract funding, it may have multiple shades of meaning depending on who is using it and where. However, there are very interesting underlying concepts concerning education behind these words: student centered learning, problem solving, multidisciplinarity and integrative education. These concepts are certainly not new: in the s XIX Dewey articulated an education system where students had to solve real life problems in a collaborative manner. But the popularization of STEAM has brought them back to the spotlight.

The integrative approach suggested by STEAM/STEM education permits solving real life problems: real problems cannot be solved just applying concepts of Maths, Physics, Technology or Engineering but utilizing a combination of all of them. In addition, it permits the integration of many different instructional methodologies such as inquiry based learning (typical from Science) and problem based learning (typical from technology/engineering).

We believe that A  (for arts and design) in STEAM is an important addition from the original STEM approach. Arts and design permit a more divergent thinking in students, giving space to more creative solutions to problems. We do not say that the convergent thinking, traditionally used in engineering or science is worse, but we believe both of them are important assets in the students toolbox. They should be able to choose which approach to use depending on the problem.

Implementing a pure STEAM education in our current education system is purely an utopia. There are plenty of challenges that teachers must cope with. For instance, they would need proper training in different fields, as well as pedagogical tools in order to be able to integrate different subjects in a single learning activity. In addition, it is necessary to overcome the disciplinary egocentrism, especially in undergraduate education. Many times, as teachers, it is difficult for us to see the links between our disciplines and others or even worse, we cannot understand how others can approach the topic from a different perspective.

For a real implementation of STEAM, especially in K-12 education, it is needed a radical change in the educational culture and administration. Curriculum and teacher assessment methods (by the administration) should change giving importance not only to the concepts but also the the acquisition of transversal competences.

This review reports several success cases of application of STEAM education in university education. We think that we should try to improve the collaboration among courses not only with colleagues in the same discipline but also in other disciplines. Organizing transdisciplinar courses (for instance, from topics from multiple faculties) might provide students more resources to face real life problems, since they would be able to face it from multiple perspectives.